\documentclass[sigconf]{acmart}
\usepackage{makecell}

\AtBeginDocument{%
  \providecommand\BibTeX{{%
    \normalfont B\kern-0.5em{\scshape i\kern-0.25em b}\kern-0.8em\TeX}}}

\setcopyright{acmcopyright}
\copyrightyear{2018}
\acmYear{2018}
\acmDOI{10.1145/1122445.1122456}

\acmConference[FDG '20]{FDG '20: ACM Foundation of Digital Games 2020 Conference}{September 15--18, 2020}{Malta}
\acmBooktitle{FDG '20: ACM Foundation of Digital Games 2020 Conference,
  September 15--18, 2020, Malta}
\acmPrice{15.00}
\acmISBN{978-1-4503-XXXX-X/18/06}

\begin{document}

\title{Open and Cultural Data Games for Learning}

\author{Domna Chiotaki}
\email{dchiotaki@gmail.com}
\affiliation{%
  \institution{University of Piraeus}
  \city{Athens, Greece}
}

\author{Kostas Karpouzis}
\email{kkarpou@cs.ntua.gr}
\orcid{0000-0002-4615-6751}
\affiliation{%
\institution{Artificial Intelligence and Learning Systems Laboratory}
\streetaddress{National Technical University of Athens}
\city{National Technical University of Athens, Greece}
\postcode{GR-157 80}
}

\renewcommand{\shortauthors}{Author 1 and Author 2}

\begin{abstract}
  Educators often seek ways to introduce gaming in the classroom in order to break the usual teaching routine, expand the usual course curriculum with additional knowledge, but mostly as a means to motivate students and increase their engagement with the course content. Even though the vast majority of students find gaming to be appealing and a welcome change to the usual teaching practice, many educators and parents doubt their educational value; in this paper, we discuss a card game designed to teach environmental matters to early elementary school students, using open data. We present a comparative study of how the game increased the students' interest for the subject, as well as their performance and engagement to the course, compared with conventional teaching and a Prezi presentation used to teach the same content to other student groups.
\end{abstract}



\maketitle

\section{Introduction}
Game-based learning (GBL) has lately been a very popular introduction to everyday school practice; a reason for this has been its relevance to 21st century skills. Game play allows students to fail with safety without any real cost. This procedure assists self-reflection and, along with the aid of critical thinking, results in strengthening the determination of the students in their several following attempts in order to succeed. In addition, creativity needed in game play triggers original thinking, inventiveness and the possibility to use failure as a way to improve and learn, along with flexibility and positivity in terms of cooperating in all situations. Games offer the necessary environment where  players-students learn to collaborate in pairs or in larger groups. Qian and Clark \cite{qian2016game} point out that the effectiveness of the Game-Based learning method relies on the design of the game. Games that have been designed based on modern theories and on successful games of the game industry are more likely to be effective in the learning of the 21st century skills.

\subsection{Games for learning}
Firstly, GBL allows students to actively involve themselves in the learning process through several stimuli. Additionally, it enables students to use previously acquired knowledge while simultaneously enriching it with new one. Apart from that, Game-Based Learning provides students with valuable feedback in real time and gives them the opportunity to self–evaluate themselves by using score or, simply, counting their victories or defeats. Finally, learning occurs empirically, by trial and error, acting and reacting or/and collaborating with fellow students. The latter also accelerates socialization \cite{oblinger2004next}.

Contrary to the traditionally accepted methods of learning, game play supports reflective thinking and drawing conclusions while, at the same time creates an environment in which experiential learning flourishes as the students are free to test and remodel their own assumptions. The games follow the ``trial and error'' approach which reinforces logical thinking and problem-solving skills \cite{vos2011effects}. In order for this method to be effective, integration of a specific lesson plan is necessary. In this plan, the traditional learning activities and the game have an equal part so as to keep student’s interest \cite{perrotta2013game}.

\subsection{Traditional learning vs Game-Based learning}
A study from MIT (Massachusetts Institute of Technology) in 2010 with a 19 years old student as a subject indicates that there is the same percentage of brain activity during watching television and attending a class \cite{guillen2012serious}. In traditional methods of learning, teachers tend to speak for an approximate 70\%-80\% of the time span of a lesson. This fact stands on the opposite side of other studies which support that a students' performance is higher when they are  more involved in the learning procedure \cite{dolan2015we}.

Tracy Sitzmann \cite{sitzmann2011meta} concluded that Game-Based Learning constitutes a better learning method as opposed to the traditional one. Specifically, self efficacy was 20\% higher after just one game, declarative knowledge was improved by 11\% and procedural knowledge by 14\%. Games, apart from being entertaining, create a competitive environment in which students immerse themselves in (engagement). On the other hand, it seems that traditional learning doesn't promote creativity and innovation through the standardized curriculum. Instead Game-Based Learning method requires creativity and initiative in order to tackle the objectives of each game through the given mechanics \cite{marsh2006reciprocal}.

\subsection{Gender and gaming}
Even though there are games which appeal at all sexes, such as Tetris \cite{cassell2000barbie}, studies indicate gender as a significant factor. Those gender-specific limitations may prevent some kids from playing with all games and, therefore, benefit from them and develop certain skills \cite{cherney2006gender}.

According to Lever \cite{lever1978sex}, boys prefer more complicated and competitive games than girls who prefer to play cooperative games. Furthermore, girls choose games with less rules and fantasy and boys those which allow the participation of multiple players. Generally, the social side of the games affects the preferences in games in both genders.

\subsection{Open data games}
Gustafsson Friberger defines \emph{data games} as ``as games where gameplay and/or game content is based on real-world data external to the game, and where gameplay supports the exploration of and learning from this data'' \cite{gustafsson2013data}. This definition is more specific in the case of \emph{open} data games. where the figures, text and facts used to populate the game are freely available for use, usually from a governmental organization or NGO. Dissemination of open data is important, since it can provide a substantial basis for argumentation in public speaking or policy making; in the context of education, it strengthens the connection between the concepts and facts taught at school with everyday life, and empowers students to make informed decisions in their life. Besides the extensive list of Open Data sources included in \cite{gustafsson2013data}, more recent, general-interest data available for reuse are available from Google's Public Data Explorer (\cite{Google_PDE}) and European Union's Open Data Portal (\cite{EU_ODP}), while FiveThirtyEight (\cite{FTE}) may be more interesting for journalists and media researchers. In the school context, however, both teachers and students are usually directed to Wikipedia and DBPedia (\cite{DBPedia}), mostly because of the friendlier, human-readable format and diversity of information.

Based on these, readily available data, a number of Data Games have emerged, from card-based games such as Open Trumps (\cite{cardona2014open}) to murder/mystery games (\cite{barros2018killed}) or creating content for existing games or other genres (\cite{barros2016playing}, \cite{barros2015balanced}). In our work, we developed a clone of the Open Trumps game, but instead of using an automated script to retrieve data from a source, we worked together with the students in order to increase their motivation and the familiarity with the game mechanics. The data we used were included in the Environmental Education course, as a part of the connection between everyday life of students and their cultural environment.

\begin{figure}[ht]
\caption{Print Top Trumps cards used to familiarize students with the rules and mechanics}
\centering
\includegraphics[width=0.5\textwidth]{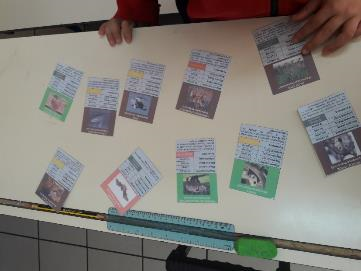}
\end{figure}

\section{Study design and implementation}
This Section discusses the aims of the study, the research questions and  hypotheses, as well as the assessment instruments used. In the first phase of the study, students filled in a pre-test questionnaire to assess their knowledge of the environmental concepts included in the cards, while the effect of the game with respect to the curriculum is measured with a similar post-test questionnaire. The group of students who played the game also filled in a post-game evaluation form about the game itself.

\subsection{Study design}
Game-based Learning has been in the limelight in recent years, both as an educational tool, as well as an ``adversary'' of learning, in the sense that gaming may distract students from the educational content. The general aim of this study is to investigate GBL (with a card game as the specific means) as a good teaching practice for environmental education, measure its effect on student performance on the subject, and also measure its impact on student interest, compared to conventional teaching. More specifically, the study aims to answer the following research questions (RQ):

\begin{itemize}
    \item \textbf{RQ1:} Investigate whether a ``top-trumps'' game can help achieve the learning goals of an environmental education course
    \item \textbf{RQ2:} Examine whether the effect of the game varies with respect to performance 
    \item \textbf{RQ3:} Study the effects of the game to students' interest in the course and its curriculum
    \item \textbf{RQ4:} Investigate any gender-related differences in performance before and after using the game
\end{itemize}

In the context of RQ1, this study aims to investigate whether the performance of the experimental group in the post-test questionnaire is correlated with the self-reported interest and fun, with the lack of anxiety or with the self-perceived ability to comprehend the concepts discussed in the course. Regarding RQ2, the study aims to identify which students in the experimental group (assessed with respect to performance) improved more than their peers, and which students in the control groups (curriculum taught via a Prezi presentation and conventional teaching, respectively) improved more, again with respect to performance tiers. For RQ3, we measured whether the performance of the experimental group increased significantly more than that of the first or the second control group, and also whether there is a significant difference between the two control groups. Finally, for RQ4, we measured performance improvements between girls and boys within the experimental group and the two control groups. In each of these questions, the null hypothesis is that there is no difference or correlation between the quantities involved (e.g. no correlation between performance and lack of anxiety in the experimental group) and the alternative hypothesis states that these quantities are correlated or differ significantly.

\subsubsection{Variables of the study}
\begin{enumerate}

\item{\textbf{Game-Based Learning}}\\
The experimental group of students used the game during the environmental education class. As a general rule, this process may lead to diverse results, with students becoming actively involved in learning and using prior knowledge in association with emerging concepts. GBL also includes instant feedback with the respect to their mastery, and the opportunity for self-assessment. Finally, learning takes place empirically, through trial and error or through collaborating with other players, also improving how students socialise during class \cite{oblinger2004next} \cite{vargianniti2019}.
As a general rule, research \cite{alsawaier2018effect} has shown that GBL increases student motivation and engagement. The best results are achieved, however, by integrating games in a specific learning process \cite{perrotta2013game}. In our case, GBL was integrated via the Top Trumps game; the outcome of this approach was compared with two other methods, that of conventional teaching and using a mixed approach, assisted with multimedia technologies (i.e. a Prezi presentation). 

\item{\textbf{Prezi presentation}}\\
Prezi \cite{Prezi} is a popular teaching tool in primary schools, mainly because it extends the usual presentation software paradigm with interactive features and easy-to-implement non-linear narrative. In our research, we used Prezi in the 1st control group and compared its results with the experimental group and with traditional teaching.

\item{\textbf{Conventional teaching}}\\
Traditionally, conventional teaching follows a teacher-centric model, with teachers transferring knowledge from school books to the students. In this context, students assume mainly a passive role, they are assessed with respect to memorizing concepts and facts and receive external rewards for their performance \cite{kordaki2011computer} \cite{Kordaki2010ACC}. Again, we used conventional teaching for the 2nd control group and compared the results to those obtained with GBL and Prezi.

\item{\textbf{Anxiety}}\\
Stress (in moderate amounts) may increase the performance of students, motivate and increase their engagement, but will impede their performance and interest when in large amounts. We measured the self-reported stress of the students \cite{baker2003prospective} \cite{park2012structural} using a questionnaire derived from the IMI (Intrinsic Motivation Inventory) \cite{goudas1994perceived} so as to investigate whether the lack of stress was correlated with performance. The questionnaire included the following questions:
\begin{itemize}
    \item{I felt no stress at all during game play}
    \item{I felt stressed during game play}
    \item{I didn't feel particularly stressed during game play}
\end{itemize}

\item{\textbf{Perception of own ability}}\\
Perception of one's ability and skill has been shown to be correlated with performance \cite{wulf1979informational} \cite{rustemeyer1984emotional}. In our research, we used an questionnaire with the following questions:
\begin{itemize}
    \item {I believe I learned a lot about mammals}
    \item {I believe I learned a lot about mammals, compared with my classmates}
    \item {I didn't do very well in the game}
\end{itemize}
in order to investigate any correlation with how students actually did in the game.

\item{\textbf{Interest - fun}}\\
We wanted to measure the fun experienced by the students, as an indicator of engagement and retention \cite{reynolds2011contrasts}. In this context, the questions we asked were again derived by the IMI test:
\begin{itemize}
    \item {I had a lot of fun playing the game}
    \item {The game was interesting}
    \item {The game was very interesting}
\end{itemize}

\item{\textbf{Gender}}\\
As a general rule, girls and boys are thought to prefer different game genres for leisure \cite{thornham2008s} and may sport different game play styles \cite{heeter2008gender}. This, however, doesn't mean that a particular game cannot be attractive and engaging for both girls and boys or that student do not play games which are popular in the other gender; overall, by not excluding games on that basis, students may improve skills which otherwise remain stagnant.

\item{\textbf{Performance}}\\
In this research, students' performance is measured with respect to the learning goals of the course and not necessarily to how well they fared in the game. We used a pre- and a post-test to estimate the effect of the each teaching approach (GBL, Prezi, conventional) to each student group.
\end{enumerate}

\subsection{Measurement instruments}
In order to measure the effectiveness of the card game with respect to the course content, as well as the student and player experience in class, we used two questionnaires. The first one (pre- and post-test) included 10 sets of questions related to the content of the course; each set included several questions with two possible answers each, and one test where the students had to list the different entries in the correct order:

\begin{enumerate}
    \item General knowledge: 2 questions about the status of conservation of different species + one ranking test
    \item Animals: 6 questions (3 easy + 3 harder) where students are asked to recognize animals shown in different images
    \item Mammals: 4 questions (2 easy + 2 harder) animal recognition questions
    \item Useful information: 6 questions about distinctive characteristics of different animals
    \item Conservation status: 4 questions (2 easy + 2 harder)  about mammals at risk. The first two questions are relatively easier, with the animals between which to choose belonging to different categories of risk
    \item Size: 4 questions (2 easy + 2 harder) comparing the size of different animals
    \item Life expectancy: 4 questions (2 easy + 2 harder) comparing the expected life span of different species
    \item Weight: 4 questions about average weight, with the first two being about animals which differ a lot
    \item Speed: again, 2 questions comparing the speed of animals with different characteristics and 2 additional ones comparing similar animals
    \item Dietary habits: 6 questions about the dietary preferences of animals
\end{enumerate}

The second questionnaire investigated how the game affected participation in class, whether it enriched the students' general knowledge about environmental issues (i.e. not directly related to the learning objectives outlined in the school curriculum) and the students' motivation. The latter part was derived from the IMI (Intrinsic Motivation Inventory) (\cite{monteiro2015intrinsic}, \cite{goudas1994perceived}) questionnaire and measured fun and interest, perception ability, and lack of anxiety or stress. All questions were answered in a 5-point Likert scale.

The game was tested with second-grade (6 y.o.) students during four teaching hours, with each group (experimental and the two control groups) consisting of 20 students. The main learning objectives included understanding of conservation status and of  measurable concepts such as speed and size, identification of different mammals, and relating the concepts taught in class with real-life example, with students acquiring both hard skills (comparing and ranking measurable concepts) and soft skills (cooperation, taking interest in environmental issues, researching open data) in the process. The first teaching session of the control group was about general concepts (herbivores, carnivores, omnivores or mammals vs egg-laying animals), wildlife preservation and animals at risk. During the second teaching session, students discussed concepts related to dietary habits and measurable concepts (size, speed, etc.), followed by a recap, a presentation and discussion of each student's favorite animal, and the post-test questionnaire; for the group using the Prezi presentation, the standard teaching material and student presentation were substituted with multimedia and interactive material.

\begin{figure}[ht]
\caption{Prezi slide showing information about herbivore animals}
\centering
\includegraphics[width=0.45\textwidth]{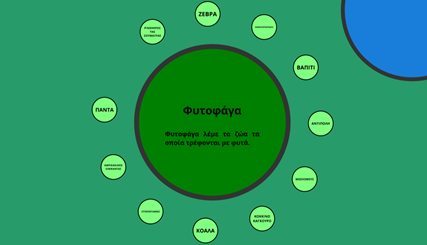}
\end{figure}

\begin{figure}[ht]
\caption{Prezi slide showing information the African elephant (\emph{Loxodonta})}
\centering
\includegraphics[width=0.45\textwidth]{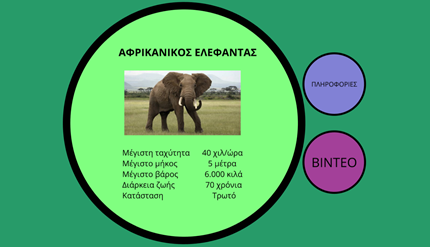}
\end{figure}

In the experimental groups, students are presented with the Top Trumps game (rules and objective) during the second teaching session and they carry on playing until the end of the third teaching session, with post-test questionnaires and presentation being completed during the last session.

\section{Results}
\subsection{Validity and reliability test}
In order to measure the validity and reliability \cite{noble2015issues} of the questionnaire, we used Cronbach's $\alpha$ \cite{bland1997statistics} to estimate its internal consistency. In order for the questionnaire to be valid, Cronbach's $\alpha$ should be greater than 0.5.

\begin{table}[ht]
\centering
\begin{tabular}{|c|c|} 
 \hline
 \textbf{Questionnaire} & \textbf{Cronbach $\alpha$} \\ 
  \hline
 1st part & 0.584 \\
  \hline
 2nd part & 0.738 \\
 \hline
 Total & 0.676 \\
 \hline
\end{tabular}
\caption{Reliability test for the measurement questionnaires}
\label{results-alpha}
\end{table}

Table \ref{results-alpha} shows that Cronbach's $\alpha$ is greater than 0.5 for both sections of the questionnaire, thus its results are consistent and valid.

\subsection{Knowledge acquisition tests}
The following table presents the results of the pre- and post-tests of each group.

\begin{table}[ht]
\centering
\begin{tabular}{|c|c|c|c|} 
 \hline
 \textbf{Group} & \makecell{\textbf{Avg.}\\\textbf{pre-test}\\\textbf{score}} & \makecell{\textbf{Avg.}\\\textbf{post-test}\\\textbf{score}} & \textbf{Difference} \\ 
  \hline
 \makecell{Experimental\\group} & 30.15 & 38.45 & 8.3 \\
  \hline
 \makecell{1st control\\group} & 29.45 & 32.25 & 2.8 \\
 \hline
 \makecell{2nd control\\group} & 30.85 & 34.15 & 3.3 \\
 \hline
\end{tabular}
\caption{Test results from pre- and post-tests for the three student groups}
\label{results-scores}
\end{table}

\begin{figure}[ht]
\caption{Pre- and post-test scores for the three student groups}
\centering
\includegraphics[width=0.45\textwidth]{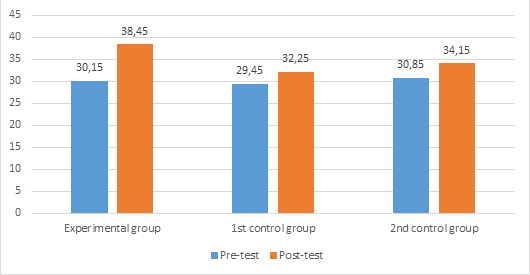}
\end{figure}

These results show that the average performance in the post-test increased by 8,3 points, i.e. 27,53\%, in the experimental group (students who used the Top Trumps games) and by 9,51\% and 10,70\% in the control groups, respectively. The following figures describe how the three groups fared in questions related to preservation status and dietary habits:

\begin{figure}[ht]
\caption{Test results in questions regarding preservation status (top) and dietary habits (bottom)}
\centering
\includegraphics[width=0.45\textwidth]{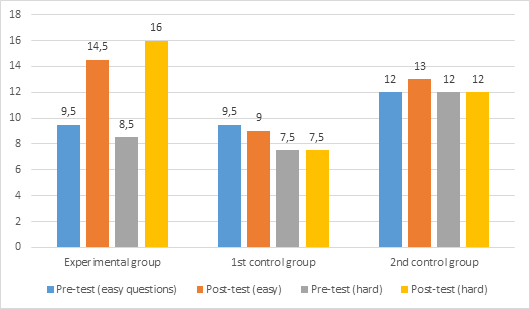}
\end{figure}

\begin{figure}[ht]

\centering
\includegraphics[width=0.45\textwidth]{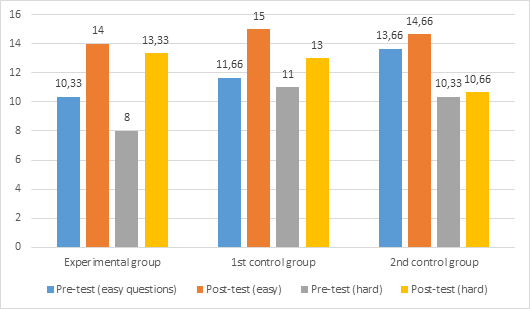}
\end{figure}

\subsection{Statistical analysis}
The second part of the questionnaire was filled in only by the students in the experimental group, since its questions were related to the game. The first question was related to how hard the game was; 55\% of the students responded with 'easy' or 'relatively easy' , which meant that the level of difficulty did not impede their performance or experience. The second question was about the means of teaching they preferred, conventional or game-based; 45\% of the students preferred game-based learning and 40\% the usual, white-board-based method. Then, students were asked about their level of stress with respect to performing well, compared to traditional teaching; 80\% of the students reported that they felt less stressed during playing the game or during the post-test. This was a welcome result for both students and teachers, since formal evaluation is also stressful for teachers, besides the usual performance anxiety on the part of the students.

The remaining entries of the questionnaire were related to fun and engagement, stress, and perception ability. Overall, 19/20 (95\%) students reported that they were interested in the game and had fun, 80\% felt no stress, and 60\% that their perception ability was increased. During observation from the researchers, students mentioned that they would like to play a Top Trumps clone game, because of its familiar mechanics and aesthetics, while the game was appealing even to students whose performance was relatively low, because of the potential to win regardless of knowing a lot about the actual subject of the game. In order to validate the results from this part of the questionnaire, we used the Shapiro-Wilk test \cite{ghasemi2012normality} to estimate the normality of the three variables.

\begin{table}[htbp]
\centering
\begin{tabular}{|c|c|c|c|c|c|c|} 
 \hline
  & \multicolumn{3}{|c|}{\makecell{Kolmogorov\\Smirnov}} & \multicolumn{3}{|c|}{\makecell{Shapiro\\Wilk}} \\
  \hline
   & \textbf{Statistic} & \textbf{df} & \textbf{Sig.} & \textbf{Statistic} & \textbf{df} & \textbf{Sig.} \\ 
  \hline
 stress & 0.195 & 20 & 0.044 & 0.868 & 20 & 0.011\\
 \hline
  interest & 0.200 & 20 & 0.035 & 0.900 & 20 & 0.041\\
 \hline
  perception & 0.177 & 20 & 0.099 & 0.875 & 20 & 0.014\\
 \hline
\end{tabular}
\caption{Tests of normality}
\label{normality-scores}
\end{table}

We also ran a normality test about the students' performance in the pre- and post-tests across all student groups. The following table shows that performance in all groups follows a normal distribution, since Sig.>0,05 for all of them.

\begin{table}[ht]
\centering
\begin{tabular}{|c|c|c|c|c|c|c|} 
 \hline
  & \multicolumn{3}{|c|}{\makecell{Kolmogorov\\Smirnov}} & \multicolumn{3}{|c|}{\makecell{Shapiro\\Wilk}} \\
  \hline
   & \textbf{Statistic} & \textbf{df} & \textbf{Sig.} & \textbf{Statistic} & \textbf{df} & \textbf{Sig.} \\ 
  \hline
 \makecell{Post-test\\(experimental)} & 0.190 & 20 & 0.057 & 0.920 & 20 & 0.101\\
 \hline
 \makecell{Post-test\\(1st group)} & 0.129 & 20 & 0.200 & 0.965 & 20 & 0.648\\
 \hline
 \makecell{Post-test\\(2nd group)} & 0.139 & 20 & 0.200 & 0.934 & 20 & 0.186\\
 \hline
 \makecell{Pre-test\\(experimental)} & 0.143 & 20 & 0.200 & 0.972 & 20 & 0.787\\
 \hline
 \makecell{Pre-test\\(1st group)} & 0.210 & 20 & 0.022 & 0.915 & 20 & 0.080\\
 \hline
 \makecell{Pre-test\\(2nd group)} & 0.090 & 20 & 0.0200 & 0.959 & 20 & 0.521\\
 \hline
\end{tabular}
\caption{Tests of normality}
\label{normality-scores-2}
\end{table}

Regarding any correlation between performance and interest and fun, lack of stress, and perception ability, we estimated Spearman's rank correlation coefficient. p-values indicated that there is no correlation between performance and fun (p=0.281>0.05), a weak positive correlation between performance and lack of stress (p=0.015<0.05) and no correlation between perception ability and performance (p=0,222>0,05). This means that the game can be fun for all students, regardless of their performance, while lack of stress is usually associated with better performance.

The next research question we examined had to do with which students (in terms of academic performance) benefit more from the game. The following figure illustrates the results.
\begin{figure}[ht]
\caption{Pre- and post-test scores for the three student groups}
\centering
\includegraphics[width=0.5\textwidth]{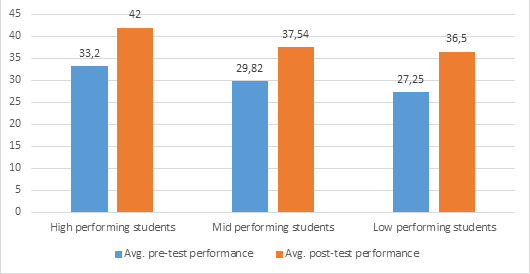}
\end{figure}

All student groups show an average increase in performance of about 8\%, with the low-performing students benefiting most from the game (9.25\%). In order to check for the statistical significance of these results, we ran a one-way ANOVA test for all three groups.

\begin{table}[ht]
\centering
\begin{tabular}{|c|c|c|c|c|c|} 
 \hline
   & \makecell{\textbf{Sum of}\\\textbf{squares}} & \textbf{df} & \textbf{Mean square} & \textbf{F} & \textbf{Sig.} \\
 \hline
  \makecell{Between\\groups} & 8.468 & 2 & 4.234 & 0.129 & 0.880 \\
 \hline
  \makecell{Within\\groups} & 559.732 & 17 & 32.925 &   &   \\
 \hline
  Total & 568.200 & 19 &   &   &   \\
 \hline
\end{tabular}
\caption{ANOVA test}
\label{anova}
\end{table}

Since, Sig.=0,880>0,05, the difference in averages is not statistically significant. This means that all students benefit equally from the game, regardless of their academic performance. Similar results (Sig.=0,198>0,05) were obtained for the control group using the Prezi presentation, while in the group taught the content using conventional teaching, better-performing students benefit less than mid- and low-performing ones. 

\section{Conclusions}
In this work, we designed a Top Trumps clone game and tested it in the framework of an Environmental Education course for early primary school students. The main research questions we wanted to investigate were whether academic performance was related to cognitive skills such as fun and interest, anxiety, and perception of ability, whether the positive or negative effect of using GBL in the classroom was different for high- or low-performing students, and whether that effect was actually positive and statistically significant; we also investigated any gender-related issues in playing the game and performing in the post-test used to measure the effect on academic performance.

Our experiment and the subsequent statistical analysis on the students' self-reported information and academic performance in the post-test showed that Game-Based Learning has a positive, statistically significant effect on academic performance in the specific field and had more impact than conventional teaching and teaching assisted by digital media. This positive effect was the same across all groups of students (with respect to performance); an interesting fact was that high-performing students did not benefit as much from conventional teaching, a fact that may be attributed to the relevant lack of interest shown by students for this paradigm. We also didn't find any statistically significant gender-related differences in game adoption and benefit from the game.

Another interesting finding had to do with the relation between interest, anxiety and performance. Since almost all students showed great or increased interest for the game, we weren't able to investigate any relation between the interest they showed and how well they did in the post-test; the lack of anxiety, though, was positively correlated with increased performance, a fact that also vouches for the introduction of more GBL sessions in the traditional curriculum.


\bibliographystyle{ACM-Reference-Format}
\bibliography{acmart}

\end{document}